\documentclass[twocolumn,showpacs,preprintnumbers,amsmath,amssymb]{revtex4}
\usepackage{graphicx}% Include figure files
\usepackage{dcolumn}% Align table columns on decimal point
\usepackage{bm}% bold math
\newcommand{\grad}{\mbox{\boldmath{$\nabla$}}}

\begin{document}

\preprint{APS/123-QED}

\title{Nonlinear Stability and Saturation of Ballooning Modes in Tokamaks}% Force line breaks with \\

\author{C.~J.~Ham}
\email{Christopher.Ham@ccfe.ac.uk}
\author{S.~C.~Cowley}
 \altaffiliation[Also at ]{Department of Physics, Imperial College, Prince Consort Road, London SW7 2BZ UK}
 \author{G.~Brochard}%
 \altaffiliation[Also at ]{Department of Physics, Imperial College, Prince Consort Road, London SW7 2BZ UK}
\affiliation{  CCFE, Culham Science Centre, Abingdon, Oxon.  OX14 3DB, UK.}

\author{H.~R.~Wilson}
\affiliation{
York Plasma Institute, Department of Physics, University of York, Heslington, York YO10 5DD UK}

\date{\today}% It is always \today, today,

\begin{abstract}
The theory of tokamak stability to nonlinear ``ballooning" displacements of elliptical magnetic flux tubes is presented.  Above a critical pressure profile the energy stored in the plasma may be lowered by finite (but not infinitesimal) displacements of such tubes (metastability). Above a higher pressure profile, the linear stability boundary, such tubes are linearly and nonlinearly unstable. The predicted saturated flux tube displacement can be of the order of the pressure gradient scale length. Plasma transport from these displaced flux tubes may explain the rapid loss of confinement in some experiments.
\end{abstract}

\pacs{Valid PACS appear here}% PACS, the Physics and Astronomy
\maketitle

%\section{\label{sec:Intro}Introduction}

Fast magnetohydrodynamic (MHD) instabilities limit the pressure (beta) in magnetically confined fusion plasmas.  The limit is observed to be one of two kinds, either a {\em soft limit} where the instability limits the pressure to a critical profile or, a {\em hard limit} where the instability rapidly destroys confinement and releases enough stored energy to take the system well below the critical pressure profile.  Sometimes the instability terminates the discharge entirely \cite{ITERdis}.  There are also two kinds of MHD instability: large scale {\em kink} instabilities and small scale, field aligned {\em ballooning} instabilities \cite{CHT}.  It is often supposed that ballooning instabilities provide a soft limit, especially near the plasma edge \cite{CTT}.  Some observations of the pressure profile evolution in the pedestal, a steep pressure gradient region at the edge of some tokamak discharges, are consistent with a soft ballooning limit \cite{WCKS, snyder}.  However {\em Edge Localised Modes} (ELMs), instabilities of the pedestal, cause an explosive eruption of multiple fine scale flux tubes and a rapid loss of edge confinement \cite{kirk}. This suggests that ballooning instabilities can sometimes provide a hard limit to edge confinement.  When then is the ballooning beta limit hard and when is it soft?

In this paper we provide a general theory of the nonlinear stability of ballooning modes.  We argue that without dissipation the nonlinear consequence of ballooning modes is the eruption of isolated elliptical magnetic flux tubes.  Certainly such elliptical erupting tubes are the long time limit of the weakly nonlinear theory developed in \cite{CA, WC}. The explosive dynamics and meta-stability of such tubes in a one dimensional line tied gravitational equilibria were studied in \cite{CCHW}.  Here we calculate the dynamics and final saturated states of erupting flux tubes; first in general (Eqs.~(\ref{infield}) and (\ref{Fperp})) and then (as an example) in a simple large aspect ratio tokamak with nearly circular flux surfaces.  The equilibrium contains a region of steep pressure gradient, a {\em transport barrier}, where the pressure gradient is of order the critical gradient for linear stability.  We adopt this equilibrium since it yields a simple nonlinear generalization of the $s - \alpha$ linear ballooning model of \cite{CHT2} and so illustrates the nonlinear dynamics. Specifically it illustrates the metastability of some linearly stable equilibria. Metastability is a phenomena encountered in many physical systems and indeed it is clear from this paper and from Ref \cite{CCHW} that many confined plasmas are also metastable. However, despite its importance, metastability in confined plasmas is largely unexplored.

\section{\label{sec:Eqbrm} Equilibrium and equations}

\begin{figure}
\includegraphics[width=0.4\textwidth]{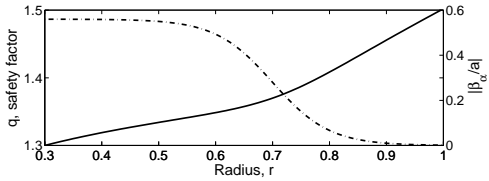}
\caption{\label{profiles}Profile of safety factor, $q$, (solid line, left hand axis) and of normalized pressure, $\beta_{\alpha}/a =2\mu_0 R_0 q^2 p_0(r)/B^2_0 a$, (dashed line, right hand axis) for the internal transport barrier (where $a$ is the plasma minor radius).} 
\end{figure}

We represent the tokamak equilibrium in flux coordinates: $\phi$ the toroidal angle, $r$ a radius like variable that is constant on a magnetic surface and $\theta$ a poloidal angle chosen to make the field lines ``straight" -- see \cite{CCHHHM, GJW}.  Thus we choose $r(\grad r\times\grad\theta) = R_0 \grad\phi$ where $R_0$ is the cylindrical radius of the magnetic axis.  Then
\begin{equation}
{\bf B}_0  =  - {\bar B}_0R_0\{ f(r)\grad r\times\grad{\cal S}\},
\label{field}\end{equation}
where $\bar{B}_0$ is a constant, ${\cal S} = \phi - q(r)(\theta - \theta_0(r))$, $q(r) $ is the safety factor and $\theta_0(r)$ is an arbitrary function of $r$.  The tokamak is large aspect ratio ({\em i.e.} $r/R_0 =\epsilon\ll 1$) and low beta $p_0(r) \sim {\cal O}(\epsilon^2 \bar{B}_0^2)$.  The {\em transport barrier} is a narrow region of steep pressure gradient ($rp_0'\sim {\cal O}(p_0/\epsilon)$) of width $\sim\epsilon r$ centred around a surface $r=r_p$ -- see Fig.~(\ref{profiles}).  The equilibrium is obtained from an expansion in $\epsilon$ (as in \cite{GJW}).

\begin{figure}
\includegraphics[width=0.4\textwidth]{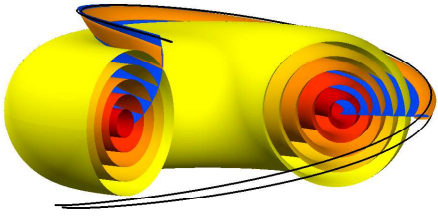}
\caption{\label{tokpic} Elliptical (orange) flux tube with $\Delta r \sim \delta_2 \gg \Delta{\cal S} \sim \delta_1$ sliding along (blue) surface ${\cal S} =0$ parting surrounding (black) field lines. Note the tube's displacement, $\xi = r - r_0$, is larger on the large $R$ part of the flux surfaces -- the tube balloons. The magnetic shear $(s=rq'/q)$ causes the twist and narrowing of the tube on the inside. } 
\end{figure}

We consider a highly elliptical flux tube of widths $\Delta r$ and $\Delta S$ with $r\gg\Delta r \sim \delta_2 \gg \Delta{\cal S} \sim \delta_1$ whose centre originates from the field line on the flux surface labelled by $r_{0}$ and ${\cal S} =0$. The field lines in the tube are displaced along the surface ${\cal S} =0$ with shape given by $r = r(\theta, r_0, t) = \xi + r_0$ where $r(t=0) = r_0$ -- see Fig.~(\ref{tokpic}). 

In principle we could consider motion along any ${\cal S}$ surface defined by any function $\theta_0(r)$ -- we restrict ourselves to the choice $\theta_0(r) = 0$.  This is the choice for the most linearly unstable motions. The tube wraps around the torus many times and we consider $r(\theta, r_0,t)$ on the domain $-\infty < \theta < \infty$. We ignore the fact that the ${\cal S} =0$ surface intersects itself as $\theta$ increases since we assume that the perturbations are sufficiently localised in $\theta$ to avoid self intersection of the flux tube.  The plasma is taken to be perfectly conducting -- {\em i.e.} the plasma is frozen to the field.  Thus the field lines must remain attached to their original surfaces and therefore $r = r(\theta, r_0, t)\rightarrow r_0$ as $|\theta|\rightarrow\infty$. The derivation of the equation of motion here follows the treatment for a general equilibrium of a magnetically confined plasma in Appendix B of \cite{CCHW2}.  The exact shape of the tube will not be needed but we do assume that $\delta_1$ is sufficiently small that we can treat the field and pressure outside the tube as unperturbed.    

We denote the field inside the tube to be ${\bf B}_{in} = {\bf B}_{in}(\theta, r_0, t)$.  The motion of the tube is assumed to be slow compared to the (sound) time to equalise pressure along the tube and thus the pressure in the tube is $p_{in}(\theta, r_0, t) = p_0(r_0)$.  The pressure forces across the tube in the direction of $\grad{\cal S}$ are formally large ($\sim p_0/\delta_1$) and therefore the total pressure inside the tube must equal the total pressure just outside the tube. Thus:
\begin{eqnarray}
B_{in}^2(\theta, r_0, t)  = B_{0}^2(\theta, r) + 2\mu_0 [p_{0}(r) - p_{0}(r_0)].
\label{infield}\end{eqnarray}
where the small perturbations of the field and pressure outside the tube are neglected (this requires $1\gg(\xi^2/R_0^2)(\delta_1^2/\delta_2^2)$). 
The ideal MHD force, $F_\perp$ pushing the field line along $\cal S$ in the direction ${\bf e}_\perp = (\grad{\cal S}\times{\bf B}_0)/B_0$ is:
\begin{eqnarray}
F_\perp =  \frac{1}{\mu_0}\left[ {\bf B}_{in}\cdot\grad{\bf B}_{in} - \grad\left(\frac{B_{in}^2}{2} + \mu_0p_{in}\right)\right]\cdot{\bf e}_\perp  \nonumber \\ 
 = \frac{1}{\mu_0}\left[ {\bf B}_{in}\cdot\grad{\bf B}_{in} - {\bf B}_{0}\cdot\grad{\bf B}_{0}\right]\cdot{\bf e}_\perp.\;\;\;\;\;\;\;\;\;
\label{Fperp}\end{eqnarray}
The second expression follows from Eq.~(\ref{infield}) and the unperturbed equilibrium relation $\grad\left(B_{0}^2/2 + \mu_0p_{0}\right) = {\bf B}_{0}\cdot\grad{\bf B}_{0}$.  Eq.~(\ref{Fperp}) is valid when the tube is sufficiently elliptical that $\delta_1^2\ll \delta_2^3/\xi$.  The expression in Eq.~(\ref{Fperp}) is a generalised form of {\em Archimedes principle} where the net force is the curvature force of the tube minus the curvature force of the tube it has displaced.  
Eqs.~(\ref{infield}) and (\ref{Fperp}) express the physics determining nonlinear ballooning -- the rest of the theory is geometry.     By requiring that ${\bf B}_{in}$ lie on ${\cal S}$ $F_\perp$ can in general be expressed in terms of $r(\theta, r_0, t)$ and its first and second derivatives with respect to $\theta$ at constant $r_0$ -- see Appendix B of \cite{CCHW2}. When $r-r_0=\xi$ is infinitesimal $F_\perp$ reduces to the familiar linear ballooning operator of \cite{CHT} -- see \cite{CCHW2}.  Note that the nonlinear force on each field line is determined independently.  The equilibrium states of the field line satisfy $F_\perp(r(\theta, r_0, t))=0$.  We model the dynamics of the tube by a simple drag evolution with ${\bf v} = v{\bf e}_\perp$, $F_\perp = \nu{\bf v}\cdot {{\bf e}_\perp}$ and $v = -R_0f\frac{\partial r}{\partial t}$. The actual dynamics of the tube are clearly more complicated but the equilibrium states must, of course, satisfy $F_\perp(r(\theta, r_0, t))=0$.    After some algebra we obtain from Eq.~(\ref{Fperp}) the evolution equation for each field line $(r(\theta, r_0, t))$ in our simple large aspect ratio model:
\begin{eqnarray}
\nu' \left(\frac{\partial r}{\partial t}\right)\left[ 1 + (\alpha\sin{\theta} - s\theta)^2 \right]  = F'_\perp(r(\theta, r_0, t))=& \nonumber \\  (\beta_{\alpha} (r_0) - \beta_{\alpha} (r))
\left[\cos{\theta} + \sin{\theta}(s\theta - \alpha\sin{\theta})\right]  &
 \nonumber \\ + 
\left(\frac{\partial}{\partial \theta}\right)_{r_0}\left(\left[ 1 + (\alpha\sin{\theta} - s\theta)^2) \right] \left(\frac{\partial r}{\partial \theta}\right)_{r_0}\right)
 \nonumber \\  - \frac{1}{2}
\left(\frac{\partial r}{\partial \theta}\right)^2_{r_0} \left(\frac{\partial}{\partial r}\right)_{\theta}(\alpha\sin{\theta} - s\theta)^2
\label{balloondrag}\end{eqnarray}
where $\nu' = \nu{\mu_0}\frac{q^2R_0^2}{B_0^2}$, $F'_\perp = F_\perp{\mu_0}\frac{qR_0^2r}{B_0^2}$, $s = rq'(r)/q(r)$ and $ \beta_{\alpha} (r) = 2{\mu_0}R_0q^2p_0(r)/{\bar B}_0^2$ and $\alpha(r) = -d\beta_{\alpha} (r)/dr$.
Eq.~(\ref{balloondrag}) is a nonlinear generalisation of the $s - \alpha$ model of \cite{CHT2}.  We define the ``energy'' functional, ${\cal{E}}(r, r_0) =\int_{-\infty}^\infty {\bf B}_{in}\cdot d{\bf r}$ where the integral is taken along the perturbed field line.\cite{CCHW2}  Note ${\cal{E}}(r, r_0)$ is formally infinite but we can make it finite by subtracting the unperturbed value  $\Delta{\cal{E}}(r,r_0) = {\cal{E}}(r,r_0) - {\cal{E}}(r_0, r_0)$. Drag evolution takes the flux tube to minima of the energy $\Delta{\cal{E}}(r,r_0)$ -- see \cite{CCHW2}.  The equilibrium states are stationary points of the variation of $\Delta{\cal{E}}(r,r_0)$ with respect to $r(\theta, r_0, t)$ at fixed $r_0$ \cite{CCHW2}.  The relative energy for our model is:
\begin{eqnarray}
\Delta{\cal{E}}(r,r_0) = \int_{-\infty}^{\infty}d\theta\left[ \frac{1}{2}
\left(\frac{\partial r}{\partial \theta}\right)^2_{r_0}\left( 1 + (\alpha\sin{\theta} - s\theta)^2 \right)
\right] - \;\;\; && \\ 
\int_{-\infty}^{\infty}d\theta\left[{\cal{A}}(r,r_0)\cos{\theta} + {\cal{B}}(r, r_0)\theta\sin{\theta} - {\cal{C}}(r, r_0)\sin^2{\theta}\right]&& \nonumber
\label{energy2}\end{eqnarray}
where the integral is at fixed $r_0$ and the energy coefficients are ${\cal{A}}(r,r_0) = \int_{r_0}^{r}(\beta_{\alpha} (r') - \beta_{\alpha} (r_0))dr'$, ${\cal{B}}(r,r_0) = \int_{r_0}^{r}(\beta_{\alpha} (r') - \beta_{\alpha} (r_0))s(r')dr'$ and ${\cal{C}}(r,r_0) = \frac{1}{2}(\beta_{\alpha} (r) - \beta_{\alpha} (r_0))^2$.

\section{\label{sec:Linear} A Linearly stable case} 

\begin{figure}
\includegraphics[width=0.4\textwidth]{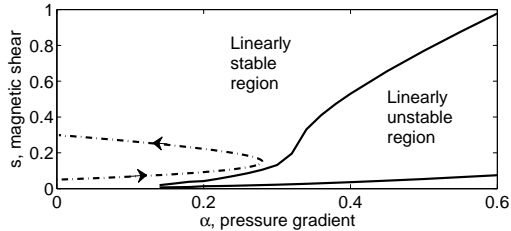}
\caption{\label{salpha} s-$\alpha$ diagram showing the linear stability boundary \cite{CHT2}.  The equilibrium here follows the trajectory of the dash-dotted line as $r_0$ is increased.  The profile is linearly stable.} 
\end{figure}

\begin{figure}
\includegraphics[width=0.4\textwidth]{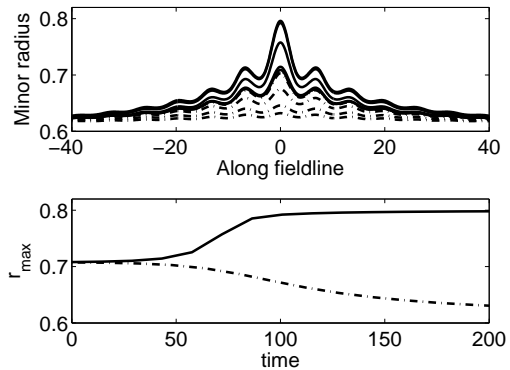}
\caption{\label{evolve2} The upper plot shows the shape of the field line at different times, $r=r(\theta,r_0,t)$ for $r_0=0.61$. The solid lines start with the initial condition just greater than the unstable equilibrium state $r_{crit}$ and evolve upwards. The dash-dotted lines start with the initial condition just less than the unstable equilibrium state $r_{crit}$ and evolve downwards. The final state of this evolution is the unperturbed field line. The lower plot shows the time evolution of maximum value along the field line $r_{max}(t)=r(0,r_0,t)$. Again, the solid (dash-dotted) line starts with the initial condition just greater (just less) than the unstable equilibrium state $r_{crit}$.} 
\end{figure}

    We investigate a case where we choose profiles of $\alpha(r)$ and $s(r)$ that yield an internal transport barrier: $\alpha(r) = \alpha_0 \textup{sech}^2\left((r-r_{\alpha})/\epsilon _p\right)$, $s(r) = (s_0 + s_1)/2+\left((s_1-s_0)/2\right) \tanh\left((r-r_s)/\epsilon _p\right)$. Linearising Eq.~(\ref{balloondrag}) with $r = \xi(\theta, r_0, t) + r_0$ with $\xi\alpha' , \xi s' \ll 1$ we obtain growing eigenmodes if the local values of $\alpha(r_0)$ and $s(r_0)$ lie in the unstable region of the s-$\alpha$ diagram  \cite{CHT2} -- see Fig.~(\ref{salpha}).  We take an initial equilibrium with no linearly unstable field lines with $\alpha_0= 0.28$, $s_0 = 0.05$, $s_1 = 0.3$, $r_a = 0.7$, $r_s = 0.72$, $ \epsilon_p = 0.1$.  As $r_0$ is increased the equilibrium traces out the dash-dotted line in Fig.~(\ref{salpha}) in the direction indicated by the arrows.  Clearly no surfaces (field lines) are linearly unstable and all infinitesimal perturbations decay.  Nonetheless finite perturbations can grow.  For example in Fig.~(\ref{evolve2}) we show the drag evolution ($r=r(\theta, r_0, t)$ using Eq.~(\ref{balloondrag})) of the field line $r_0 = 0.61$ with two finite initial displacements.  
%\begin{figure}
%\includegraphics[width=0.4\textwidth]{EvolveUP}
%\caption{\label{evolveup} The surface illustrates the drag evolution of $r = r(\theta, r_0, t)$ for $r_0=0.61$ with an initial displacement larger than the critical perturbation.  The final state of this evolution is the stable displaced equilibrium -- essentially the $t>100$ line.  The critical perturbation, the unstable displaced equilibrium, is given by the $t=0$ line.} 
%\end{figure} 
%\begin{figure}
%\includegraphics[width=0.4\textwidth]{EvolveDN}
%\caption{\label{evolvedn} The surface illustrates the drag evolution of $r = r(\theta, r_0, t)$ for $r_0=0.61$ with an initial displacement smaller than the critical perturbation. The final state of this evolution is the unperturbed field line.} 
%\end{figure} 
The larger initial displacement evolves to a finite displaced stable equilibrium. The smaller initial displacement decays to the linearly stable unperturbed state $r=r_0$ (Fig.~(\ref{evolve2})).  There are three equilibrium states of this field line that can be found by solving the equation $F'_\perp =0$ (see Eq.~(\ref{balloondrag})) by a simple shooting method.  These are: the linearly stable unperturbed state $r=r_0$ with relative energy $\Delta{\cal{E}}=0$; an unstable equilibrium state, $r=r_{crit}(\theta,r_0)$, between the two initial conditions shown at $t=0$ in Fig.~(\ref{evolve2}) with $\Delta{\cal{E}}=1.09\times10^{-4}$ and; the stable equilibrium state, $r=r_{sat}(\theta,r_0)$ that is the final state of the larger perturbation with $\Delta{\cal{E}}=-0.8\times10^{-4}$.  Clearly the unperturbed state is meta-stable since a finite perturbation triggers evolution to a lower energy state.

\begin{figure}
\includegraphics[width=0.4\textwidth]{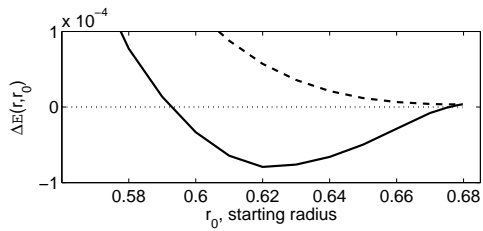}
\caption{\label{energyplot} . Relative energy, $\Delta{\cal E}$, evaluated from Eq.(\ref{energy2}) for three equilibrium solutions, $F'_\perp =0$, of Eq.(\ref{balloondrag}) in the region $0.58<r_0 <0.68$.  The dotted line is the unperturbed energy, the dashed line is the unstable displaced equilibrium energy ($\Delta {\cal E}(r_{crit},r_0)$) and the solid line is the displaced stable equilibrium energy ($\Delta {\cal E}(r_{sat},r_0)$). The stable displaced equilibrium is the lowest energy state for $0.593< r_0 < 0.679$.} 
\end{figure}

\begin{figure}
\includegraphics[width=0.4\textwidth]{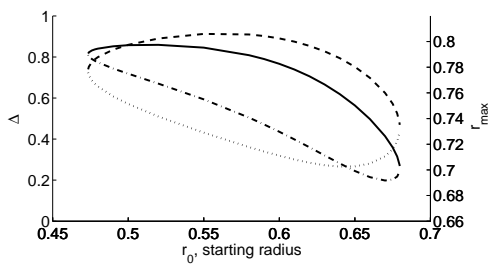}
\caption{\label{dbeta} A measure of the ballooning displacement $\Delta = (\beta_{\alpha} (r_0) - \beta_{\alpha} (r_{max})/(2\epsilon _p \alpha_0)$ for the two perturbed equilibrium states (left-hand axis).  Field lines in the displaced lower energy equilibrium can cross a substantial fraction of the pressure profile (solid line) -- for example the $r_0 = 0.61$ field line balloons across about $73\%$ of the pressure profile. The unstable equilibrium is shown by the dash-dotted line. The $r_{max}$ for the saturated field lines is shown as the dashed line (right-hand axis) and the $r_{max}$ for the unstable equilibria are shown as the dotted line. Note that for $0.56<r_0<0.68$ the field lines ``overtake'' i.e. $r_{max}(r^1_0)> r_{max}(r^2_0)$ if $r^1_0<r^2_0$. } 
\end{figure}

Not all the field lines have lower energy equilibrium states. We have examined the $F'_\perp =0$ solutions for $0.4< r_0<0.8$. For $0.474< r_0 < 0.680$ there are three equilibrium solutions but outside this region the only equilibrium solution is the unperturbed state.  All displaced solutions are even in $\theta$ and have their maximum displacement at $\theta=0$ which we denote $r_{max}$.  In Fig.~(\ref{energyplot}) we plot $\Delta{\cal{E}}$ for $0.58<r_0 <0.68$ and in Fig.~(\ref{dbeta}) we plot both $\Delta = (\beta_{\alpha} (r_0) - \beta_{\alpha} (r_{max})/(2\epsilon _p \alpha_0)$ (solid and dash-dotted lines, left-hand axis) and $r_{max}$ (dashed and dotted lines, right-hand axis).  $\Delta$ measures the fraction of pressure profile crossed by the ballooning flux tube.  Clearly for $0.593< r_0 < 0.678$ the lowest energy state is a displaced state (the solid black line in Fig.~(\ref{energyplot})) -- these states can be reached by giving the field line a perturbation with more than the energy of the unstable positive energy equilibrium state (the dashed line in Fig.~(\ref{energyplot}))


We varied $\alpha_0$ for this case; for $\alpha_0 \leq 0.269$ there are no energetically favourable saturated states, and for $\alpha_0 \geq 0.306$ some field lines are linearly unstable. It is not however the linearly unstable field lines that produce the saturated field lines with the largest displacement. These are metastable field lines with $r_0 \approx 0.6$. For a linearly unstable field line there are two lower energy saturated states, one displaced outwards and one inwards.

\section{\label{sec:Discussion} Discussion}

In this paper we have formulated a nonlinear theory of ballooning modes as the eruption of elliptical flux tubes.  The force in the direction of motion of the flux tube is given by combining pressure balance across the elliptical tube (Eq.~(\ref{infield})) with a generalised Archimedes principle (Eq.~(\ref{Fperp})).  We illustrate our theory with the drag evolution of flux tubes in a large aspect ratio circular flux surface equilibrium with an {\em internal transport barrier} -- a nonlinear $s-\alpha$ model \cite{CHT2}.
This model reveals remarkable physics.  Even below the linear stability threshold there can be lower energy saturated flux tubes with finite displacement -- we have found such states, see Fig.~(\ref{energyplot}).

%The drag evolution model that we have used to produce the time dependent shape of the flux tube is an approximation to the real non-linear dynamics of the flux tube. However, this model is likely to capture the key features, such as the explosive nature of the evolution which is seen in Fig.~(\ref{evolve2})). The flux tube has been assumed to have an elliptical shape in this calculation. This is based on the shape of the mildly nonlinear flux tubes \cite{CCHW, CA, WC} and physical intuition. Specifically the elliptical shape minimizes the stabilizing sideway motion of field lines outside the erupting flux tube.  Overtaking may alter the assumed elliptical flux tube cross section (see Fig. (\ref{dbeta})). Further work is needed to understand the cross section shape and its evolution.

The flux tubes have been modelled with a perfectly conducting plasma. This is a reasonable assumption since the eruption is likely to take place on a fast timescale. Once the flux tubes have reached their saturated states then other, slower timescale, processes will become important. For example resistive field line reconnection is likely to occur at large $\theta$ as it does in resistive ballooning modes \cite{Strauss}. There is also likely to be cross field transport of heat from the tube to the surrounding plasma around $r_{max}$ given the large gradient of temperature. This would effectively connect the high pressure region to the low pressure region via a conduit (``hosepipe") along the flux tube -- perhaps causing rapid loss of confinement locally. The balance of the dissipative processes will determine the longer timescale evolution of the flux tube and ultimately how it disconnects from or returns to, its original location.

ELMs are a possible application of the ideas in this paper. However, we leave this topic to a future publication. We instead note that the explosive eruption of ballooning modes have been observed in TFTR shots with internal transport barriers \cite{Fredrickson}. A slowly evolving $n=1$ kink mode arises first and then a toroidally localized ballooning mode (with $n\sim 10-20$) appears. These ballooning modes eventually disrupt the plasma -- a hard limit. The tubes could be destabilised from a meta-stable state at the tip of the kink mode by finite noise or by passing through linear marginal stability.  We have demonstrated with the model above that flux tubes can erupt into finitely displaced states effectively connecting plasma inside the transport barrier to outside the barrier.  We speculate that the ballooning mode provides a hard limit when and only when there are {\em finitely} displaced lower energy saturated states.  However there is, clearly, much to understand before we can claim to fully understand the hard/soft limit of ballooning modes.             

This work has received funding from the European Union's Horizon 2020 research and innovation programme, grant 633053, and from the RCUK Energy Programme [grant EP/I501045]. For further information on this paper contact PublicationsManager@ccfe.ac.uk. The views and opinions expressed herein do not necessarily reflect those of the European Commission. H.R. Wilson is a Royal Society Wolfson Research Merit Award holder. 

%\newpage %Just because of unusual number of tables stacked at end

\begin{thebibliography}{99}
\bibitem{ITERdis}  T.C. Hender, et. al.  Nucl.\ Fusion, {\bf 47}, S128 - S202(2007).
\bibitem{CHT} J.\ W.\ Connor, R.\ J.\ Hastie and J.\ B.\ Taylor, Proc.\ R.\ Soc.\ London {\bf A365}  (1979)
\bibitem{CTT} J.\ W.\ Connor, J.\ B.\ Taylor, and M.\ Turner, Nucl.\ Fusion {\bf 24} 642(1984)
\bibitem{WCKS} H.\ R.\ Wilson, S.\ C.\ Cowley, A.\ Kirk and P.\ B.\ Snyder { \it Plasma Phys. Control. Fusion} {\bf 48} A71-A84 (2006)  
\bibitem{snyder} P.\ B.\ Snyder, et. al., Nucl.\ Fusion, {\bf 51}, 103016 (2011).
\bibitem{kirk} A.\ Kirk , et. al., Phys.\ Rev.\ Lett. {\bf 96}  185001 (2006)
\bibitem{CA}S. C. Cowley and M. Artun, {\it Physics Reports,} vol. 283, pp. 185 -- 211, 1997. 
\bibitem{WC} H.\ R.\ Wilson and S.\ C.\ Cowley, {\it Phys. Rev. Lett.},  vol. 92,  no. 17, 175006-1-- 175006-4, 2004.
\bibitem{CCHW} S.\ C.\ Cowley, B.\ Cowley, S.\ A.\ Henneberg and H.\ R.\ Wilson,  Proc. R. Soc. A {\bf 471}  20140913(2015), 
\bibitem{CHT2} J.\ W.\ Connor, R.\ J.\ Hastie and J.\ B.\ Taylor, Phys.\ Rev.\ Lett. {\bf 40}  396(1978)
\bibitem{CCHHHM} J.\ W.\ Connor, et. al., Phys.\ Fluids. {\bf 31}  577(1988)
\bibitem{GJW} J.\ M.\ Greene, J.\ L.\ Johnson and K.\ E.\ Weimer, Phys.\ Fluids. {\bf 14}  671 (1971)
\bibitem{CCHW2} S.\ C.\ Cowley, B.\ Cowley, S.\ A.\ Henneberg and H.\ R.\ Wilson,  arXiv preprint arXiv:1411.7797 (2014), 
%\bibitem{Wolf} R.\ C.\ Wolf { \it Plasma Phys. Control. Fusion} {\bf 45} R1 (2003)
\bibitem{Strauss} H.\ R.\ Struass { \it Phys. Fluids } {\bf 24} 2004 (1981)
\bibitem{Fredrickson} E.\ D.\ Fredrickson, et. al. { \it Phys. Plasmas } {\bf 3} 2620 (1996)
\end{thebibliography}

\end{document}